\documentclass[%
 reprint,
 amsmath,amssymb,
 aps,
]{revtex4-2}

\usepackage{graphicx}
\usepackage{dcolumn}
\usepackage{bm}
\usepackage{multirow}
\usepackage{lineno}

\usepackage{float}

\usepackage{url}

\begin{document}


\title{Evaluating the root causes of fatigue and associated risk factors in the Brazilian regular aviation industry}

\author{Tulio E. Rodrigues$^{1,2,*}$, Frida M. Fischer$^{3} $, Otaviano Helene$^{1}$, Eduardo Antunes$^{4}$, Eduardo Furlan$^{2}$, Eduardo Morteo$^{2}$, Alfredo Menquini$^{5}$, João Lisboa$^{4}$, Arnaldo Frank$^{5}$, Alexandre Simões$^{6}$, Karla Papazian$^{7}$ and André F. Helene$^{7}$ \vspace{0.15in}}

\affiliation{$^{1}$Experimental Physics Department, Physics Institute, University of São Paulo, P. O. Box 66318, CEP 05315-970, São Paulo,
Brazil}
\affiliation{$^{2}$ Technical Board, Gol Aircrew Association (ASAGOL), São Paulo, Brazil}
\affiliation{$^{3}$ Department of Environmental Health, School of Public Health, University of São Paulo, São Paulo, Brazil}
\affiliation{$^{4}$Flight Safety Board, National Aircrew Union (SNA), São Paulo, Brazil}
\affiliation{$^{5}$Technical Board, Brazilian Association of Civil Aviation Pilots (ABRAPAC), São Paulo, Brazil}
\affiliation{$^{6}$Safety Board, LATAM Aircrew Association (ATL), São Paulo, Brazil}
\affiliation{$^{7}$Department of Physiology, Institute of Biosciences, University of São Paulo, São Paulo, Brazil\vspace{0.15in}}

\date{\today}

\begin{abstract}
This work evaluates the potential root causes of fatigue and its relationships with accident risks using a biomathematical model approach and a robust sample (N = 8476) of aircrew rosters from the Brazilian regular aviation, extracted from the \textit{Fadigômetro} database. The fatigue outcomes derive from the software Sleep, Activity, Fatigue, and Task Effectiveness Fatigue Avoidance Scheduling Tool (SAFTE-FAST), which considers the homeostatic process, circadian rhythms and the sleep inertia. The analyses include data from January 2019 until March 2020 and show relevant group effects comparing early 2019 and 2020, with the latter presenting lower fatigue outcomes in most cases. The average minimum SAFTE-FAST effectiveness during critical phases of flight (departures and landings) decreases cubically with the number of shifts that elapse totally or partially between mid-night and 6 a.m. within a 30-day period ($ N_{NS}$). As a consequence, the relative fatigue risk increases by 23.3\% (95\% CI, 20.4-26.2\%) when increasing $ N_{NS}$ from 1 to 13. The average maximum equivalent wakefulness in critical phases also increases cubically with the number of night shifts and exceeds 24 hours for rosters with $ N_{NS}$ above 10. The average fatigue hazard area in critical phases of flight varies quadratically with the number of departures and landings within 2 and 6 a.m. ($ N_{Wocl}$). These findings demonstrate that both $ N_{NS}$ and $ N_{Wocl}$ should be considered as key performance indicators and be kept as low as reasonably practical when building aircrew rosters, in order to properly manage the fatigue risk. All the fittings were performed using the Least Square Method and the confidence intervals were calculated using uncertainty propagation techniques and the covariance matrix of the fitted parameters. The effectiveness scores obtained at 30 minute time intervals allowed a model estimate for the relative fatigue risk as a function of the time of the day, whose averaged values show reasonable qualitative agreement with previous measurements of pilot errors in the cockpit. Moreover, the 2019 data revealed a risk exposure factor two times (14\%) higher than the figures reported by Mello et al. (2008) (7\%), within 0h00 do 05h59. Tailored analyses of the SAFTE-FAST inputs for afternoon naps before night shifts, commuting from home to station and vice-versa, and bedtime before early-start shifts were carried out using the responses of a questionnaire. Wilcoxon signed-rank tests for matched pairs show significant group effects ($ p < 0.001 $) comparing the groups with and without afternoon naps, with one or two hours of commuting and with or without the advanced bedtime feature of the SAFTE-FAST software. The average fatigue hazard area in critical phases of flight increases by 43 to 63\% switching off the afternoon naps, 14 to 21\% increasing the commuting from one to two hours and 35 to 54\% switching off the advanced bedtime criterion, evidencing the need of a better and more accurate understanding of these parameters when modelling fatigue risk factors.

\end{abstract}


\email{tulio@if.usp.br}

\maketitle


\section{Introduction}

Until the Covid-19 outbreak since early 2020, the commercial aviation industry has experienced a solid growth during the last decades and it is particularly characterized as a high standard, strongly regulated and safe transport category \citep{ICAO2020, janic2000assessment}. As airplane systems became more reliable, special attention has been given for human errors, which, according to a Boeing summary \citep{boeing2001}, have contributed with 66\% of commercial jet fleet hull-loss accidents between 1992 and 2000.
Among all the relevant aspects related with human factors \citep{kharoufah2018review}, the physiological issue of mental fatigue plays a relevant role in a 24/7 society \citep{caldwell2005}.
In this regard, countries worldwide are making efforts to establish effective barriers to mitigate the fatigue risk either via prescriptive flight and duty time limitations or via the implementation of fatigue risk management systems \citep{Icao2016}.
A recent experiment carried out by the European Aviation Safety Agency (EASA) \citep{EASA2019} investigated the effectiveness of some prescriptive rules and scenarios in Europe using two well-known biomathematical models: the Sleep, Activity, Fatigue, and Task Effectiveness Fatigue Avoidance Scheduling Tool (SAFTE-FAST) \citep{Hur2004} and the Boeing Alertness Model (BAM) \citep{ingre2014}. One of their findings revealed that despite of being necessary, the prescriptive rules are not fully sufficient to mitigate the fatigue risks, specially during disruptive and/or night shifts. Moreover, the Working Time Society has recently stated that prescriptive rules ignore biological aspects and become less effective for working activities outside the normal daytime hours \citep{honn2019}. Such findings reinforce the importance of a solid safety culture in organizations, as well as the need to strengthen airlines' safety management systems (SMS), bridging the gap between the industry needs and the scientific knowledge in the field.\\
In Brazil, a new set of prescriptive limits and labour clauses was put in place by August 2017 in law 13.475/17 \citep{Brasil2017}. After this change, the National Civil Aviation Agency (ANAC) established the criteria and requirements for the implementation by the airlines of a fatigue risk management program/system based on international civil aviation standards and recommendations \citep{Icao2016} via the Brazilian Civil Aviation Regulation (RBAC 117) \citep{Anac2019}. Such set of rules from RBAC 117 allows some extensions of the prescriptive limits dictated by the law 13.475/17, as far as the operators comply with additional requirements.\\
Together with the global efforts of managing fatigue risks, guidelines specially developed for the identification of fatigue as a contributing factor in aviation accidents and serious incidents, like the one recently published by the National Commission of Human Fatigue \citep{CNFH2020} (a Brazilian Commission composed by several stakeholders), also improves the overall knowledge of the likely causes of fatigue, driving future preventive measures and improvements for the aviation system. Moreover, the analysis of speech parameters and its correlations with fatigue and sleepiness under operational circumstances is also a good practical example of research being successfully applied to aviation accident investigations \citep{de2019speech}.\\ 
Several model based approaches have been used to predict fatigue and/or sleepiness outcomes due to sleep loss and/or circadian disruptions \citep{Hur2004, roma2012, ingre2014, ran2013, raslear2011}. More recently, Cochrane \textit{et al}. also emphasized the importance of considering non-linear relationships between fatigue and risk \citep{cochrane2021ensemble}. In a simulated space mission with sleep restrictions, Flynn‑Evans \textit{et al}. successfully described average changes in performance with bio-mathematical models \citep{flynn2020changes}. Reviews of the basic features of some of the available biomathematical models usually adopted in the aviation industry can be found elsewhere \citep{mallis2004, van2004, CASA2014}.\\
In Brazil, a study carried out in 2012 found strong evidences of a chronic fatigue scenario by the analysis of pilot reports using the SAFTE-FAST model \citep{licati2015}. Another study, derived from the continued fatigue monitoring effort named \textit{Fadigômetro}, has shown relevant seasonal variations in fatigue indicators comparing high and low season rosters of 2018 \citep{rod2020}. This first finding demonstrates the potentialities of the \textit{Fadigômetro} project as a reliable tool for the analysis of the impact of regulatory changes, such as the one in effect since March 2020, when the new rules set by RBAC 117 became effective. Unfortunately, the outbreak of Covid-19 in Brazil by the second half of March 2020 coincided with the regulatory change, postponing an unbiased analysis of its impact until the aviation fully recovers \citep{Iba2021}.\\
In this work, we present a detailed statistical analysis, based on the SAFTE-FAST model outputs and other key performance indicators, for the investigation of the root causes of fatigue and its corresponding relative fatigue risks in a huge sample of rosters of the Brazilian regular aviation, extracted from the \textit{Fadigômetro} database.

\section{Methods}
\subsection{The Sample}

This work included 8476 executed rosters from January 2019 up to mid-March 2020 of aircrew workers pertaining to major Brazilian airlines and a questionnaire, all extracted from the \textit{Fadigômetro} database on March 2, 2021. The period of analysis was chosen with the aim of capturing fatigue outcomes in a recent past scenario before the Covid-19 outbreak in Brazil (mid-March of 2020).\\
The questionnaire included sociodemographic, behaviour and health questions and was filled up by 796 participants. The eligible airlines altogether comprised 92.5\% of the Brazilian regular aviation market share in 2019 \citep{ANAC2020}.\\

\subsection{Ethical considerations}
The present work is derived from the project "Analysis of Fatigue in Brazilian Civil Aviation" approved by the research ethics committee of the Institute of Biosciences, University of São Paulo (Certificate of Presentation for Ethical Appraisal no. 89058318.7.0000.5464). It was ensured confidentiality to all eligible subjects who voluntarily agreed to participate by approving a digital informed consent form. We declare no commercial, labor duality or conflict of interest with any representative institution involved in the experiment, airline or regulator. Confidentiality was ensured for the airlines whose rosters were analysed.

\subsection{Rosters: criteria and filtering}
As described elsewhere \citep{rod2020} rosters were automatically fed into a web-based application, being digitized and analysed by means of an on-line algorithm. Some filters were implemented to extract the comma-separated values (CSV) files, including an internal identification number (Id), event type (\textit{Crewing} for flights and \textit{Working} for non-crewing tasks), departure/landing times and locations, crew rank, contractual basis and start/end of the duty times. Differently from our previous work \citep{rod2020}, which included only \textit{Crewing} events, this study also considers all \textit{Working} events in the rosters, such as standbys, training activities, flying as a passenger for airline purposes (dead-heading flights), etc. These non-crewing events do not directly contribute for risk build-up, but may adversely interfere on the sleep opportunities of the subjects, which, in turn, affects their alertness levels and the overall fatigue and/or sleepiness outcomes. The inclusion of all the \textit{Working} events in the analyses increased the occurrences of inconsistencies in the extracted CSV files, most frequently associated with erroneous/spurious information, which caused crashes in the SAFTE-FAST (SF) runs. Furthermore, few rosters also presented warnings in the SF console due to the large majority of events being associated with on ground training activities. Among all the eligible rosters, 0.64\% to 4.5\% had crashes, and 0.2\% to 6.3\% warnings. All rosters with problems were excluded from the analyses, resulting in a total of 8476 validated rosters.\\
Considering that only executed rosters were included (past events), home standbys - which are \textit{Working} activities where the crew member stays on-call at the place of their choice - were disregarded in the analyses. Such procedure avoids any bias from the model assumption that no sleep event will occur during these working periods, which does not seem very likely in a realistic scenario particularly for home standbys during night. As a consequence of this criterion, our results should be interpreted as lower fatigue limits, as some of the participants might have poor or actually no sleep while in home standbys during night, despite of being, for instance, at home.\\
Similarly as described elsewhere \citep{rod2020}, additional filters of minimum crew and narrow body aircraft were also applied in order to focus on the Brazilian domestic flights and few mid-haul international flights within Caribbean, South and North America executed by minimum crew. Such choice prevents the inclusion of augmented crew flights, which are characterized by inboard sleep opportunities, a feature not included in our SF input criteria for this analysis (see Supplementary Section). Since some of the key performance metrics, such as the fatigue hazard area (\textit{FHA}), strictly depend on the time interval of the analysis, epochs of exactly 30 days were fixed for each month (see Supplementary Section).\\

\subsection{Modelling the fatigue risk} 
 
Following the steps described in our previous work \cite{rod2020}, some key performance indicators, such as the SF Effectiveness ($E_{SF}$), are useful to address the probability of mental fatigue and its relationship with the risk of cognitive impairment, mishaps and, ultimately, serious incidents or accidents. The SF model \citep{Hur2004} is a three step algorithm that takes into account the homeostatic process, the circadian rhythms associated with sleep and wakefulness and the sleep inertia. Such biomatematical model has being successfully validated against human factor railroad accidents \citep{hur2006, hur2011}, as well as with psychomotor vigilance test (PVT) measurements under aviation operational environments \citep{roma2012} and in a simulated space mission with sleep restrictions \citep{flynn2020changes}. As we have shown recently, the relative probability of railroad accidents caused by human factors, herein denoted as $P_{HF}$, increases inversely with $E_{SF}$, such that:
\begin{equation}
\label{eqn:1}
P_{HF}(E_{SF})=b/E_{SF}, 
\end{equation}
with $ b = 79.6\pm3.0 \%$ and $E_{SF}$ given as a percentage from 0 to 100\%, where 100\% represents an optimum individual performance \cite{rod2020}.\\
During the so-called Window of Circadian Low (WOCL), which is considered as default from 2 to 6 a.m. in the SF model, individuals have a higher probability of being fatigued and/or sleepy, with usually lower performance scores related with alertness and attention. However, in the 24/7 aviation industry, flight operations are still needed within these less favourable hours of the day, requiring additional protective barriers to mitigate the fatigue risks. In this regard, it is very useful to investigate the $E_{SF}$ scores in the critical phases of flight, which comprise the first and last 30 minutes of each flight sector \cite{rod2020}. Consequently, the most degraded fatigue scores, usually called hot spots of fatigue, for a given crew-member within a period of analysis can be given by the minimum $E_{SF}$ score in the critical phases of flight for \textit{Crewing} events, herein denoted as $EM_{C}$. Additionally with $EM_{C}$, the minimum sleep reservoir (\textit{R}) in the critical phases of flight, herein denoted as $ RM_{C}$, also represents a key performance variable strictly related with the sleep debt ($ SD $) and wakefulness ($t_{awake} $). As described elsewhere \citep{Hur2004}, the sleep reservoir \textit{R} varies from 0 to 100\%, increasing during sleep periods and decreasing during wakefulness, with a score of 75\% representing 8 hours of sleep debt. So, considering the equations presented in Ref. \citep{Hur2004}, the following linear relationships hold:
\begin{equation}
\label{eqn:2}
SD=32\left( 1-\frac{R}{100}\right) 
\end{equation}
\begin{center}
and
\end{center}
\begin{equation}
\label{eqn:3}
t_{awake}=\frac{2880}{0.5\times60}\left( 1-\frac{R}{100}\right)= 3\times SD,
\end{equation}
with $SD$ and $t_{awake} $ in hours. So, given a minimum sleep reservoir in critical phases of flight one can easily obtain the corresponding maximum sleep debt $SD^{max}$ and the maximum equivalent time awake $t_{awake}^{max}$. \\      
Differently from $EM_{C}$ and $ RM_{C}$, which represent plausible metrics to identify fatigue hot spots in crew rosters, the \textit{FHA} brings the concept of a cumulative fatigue score for a given individual within a given period of analysis. This metric was first proposed by Rangan and Van Dongen in 2013 \citep{ran2013} and represents the area of the SF effectiveness lineshape along time under a given threshold. Such additive metric represents an overall quantitative fatigue score, which could help to guide preventive actions to mitigate the fatigue risks in hundreds or even thousands of crew rosters. Following the same steps described elsewhere \citep{rod2020}, we have adopted a SF effectiveness threshold of 77\% in order to calculate the \textit{FHA} during critical phases of flight, herein denoted as $FHA_{C}$.\\
The SF parameters and criteria used in our model calculations are described in detail in the Supplementary Section and are strictly the same adopted in our previous work \citep{rod2020}. However, in this work, we have also investigated - using behavioural information from the questionnaire - three important SF inputs: (1) afternoon naps prior to night shifts, (2) commuting from home to station and vice-versa, and (3) the advanced bedtime feature of the software. For the afternoon naps prior to night shifts, the standard parametrization of the software considers no nap for the individuals with less than 8 hours since the last sleep event and a 60, 90, 120 and 180 minute nap if the individual is within 8 to 10 hours, 10 to 12 hours, 12 to 14 hours or more than 14 hours since the last sleep event, respectively. For this input a tailored analysis was carried out switching off the Auto-Nap feature, which means the software will not include afternoon naps regardless of the wakefulness period, for those individuals who declared not being used to take any nap before night shifts [364 out of 796 responders (45.6\%)]. For the commuting from home to station and vice-versa, we have also run analyses with two hours (\textit{extended} commuting), in contrast with our \textit{standard} parametrization of one hour. For this input, we have considered the individuals that declared a commuting of two hours [200 out of 796 responders (25.1\%)]. For the advanced bedtime feature, the software assumes that individuals go to bed earlier than usual (considering the standard bedtime of 11 p.m. \citep{rod2020}), as a sleep strategy before early-start shifts, typically between 6 and 8 a.m., regardless of having or not any significant sleep deprivation prior to the sleep event. So, in our calculations, we have investigated a scenario without the advanced bedtime feature of the SF model for the individuals who reported not doing any anticipation of the bedtime [262 out of 796 of the responders (31.9\%)]. All these customized runs were performed for two high (February and July) and two low (May and June) productivity months of 2019.

\subsection{Variables, statistical analyses and fitting procedures}    

The dependent variables include: $EM_{C}$, $RM_{C}$, $FHA_{C}$, $SD^{max}$, $t_{awake}^{max}$ and $P_{HF}$. 
The independent variables include: time of the day, $ t_{clock}$; duty time, $DT$; number of night shifts, $N_{NS}$; number of consecutive night shifts, $N_{CNS}$; number of \textit{Working} events, $N_{work}$; number of \textit{Crewing} events (flight sectors), $N_{crew}$; and number of WOCL events, $N_{wocl}$. Both $N_{NS}$ and $N_{CNS}$ include \textit{Crewing} and \textit{Working} events where any portion of the duty period occurs between mid-night and 6 a.m. $N_{wocl}$ includes all the departures and arrivals within 2 and 6 a.m. for \textit{Crewing} events only. The independent variables $DT$, $N_{NS}$, $N_{CNS}$, $N_{work}$, $N_{crew}$ and $N_{wocl}$ are closely related with workload and will be denoted throughout the paper as \textit{productivity metrics}. All the analyses were done for exact 30-day epochs to avoid any bias when comparing months of different lengths (see Supplementary Section for details).
Another set of dependent variables derived from our calculations include the relative fatigue risk as a function of $N_{NS}$, $RFR(N_{NS})$; the SF effectiveness as a function of $t_{clock}$, $E_{SF}(t_{clock})$; the relative fatigue risk as a function of $t_{clock}$, $RFR(t_{clock})$; and the flight proportion as a function of $t_{clock}$, $FP(t_{clock})$.\\
For the normality hypothesis we have adopted the Shapiro-Wilk test \citep{Sha1965}. For the evaluation of group effects we applied the Mann-Whitney test for two independent samples and the Wilcoxon signed-rank test for paired samples, depending on the situation. All the statistical tests were performed with the IBM SPSS software version 25. The calculations of the effect size (dz) were performed by G*Power version 3.1.9.7 \citep{faul2007g}.\\
All the fitting procedures were carried out using the least square method \citep{helene2016useful}, where the best fit parameters correspond to the minimum $ \chi^{2}$, defined as:
\begin{equation}
\label{eqn:4}
\chi^2=\sum_{i=1}^{n}\frac{[ \tilde{f}(x_{i}) - y_{i}] ^{2}}{\sigma y_{i}^{2}},
\end{equation}
where $\tilde{f}(x_{i})$ stands for the fitted function calculated at each $ x_{i}$ value, $ y_{i} $ the corresponding data point, $\sigma y_{i}$ its respective standard error and \textit{n} the total number of data points to be fitted. Given that all the data presented throughout this paper depend linearly on the parameters, the optimal properties of the least square method of minimum variance and unbiased fitting are fully satisfied. All the fits are considered successful if the probability of exceeding the $\chi^2$ (\textit{p}-value) is $\geq 0.05.$\\
The uncertainties of the fitted functions $\sigma_{\tilde{f}}$ where obtained by the propagation of the uncertainties of the fitted parameters taking into account its full covariance matrix, as similarly described in a recent calculation applied for the COVID-19 pandemic spread \citep{rodrigues2020monte}. The 95\% confidence intervals of the fitted functions were assumed as $\sim 2\sigma_{\tilde{f}} $.
In some model estimates, standard uncertainty propagation techniques were also applied \citep{helene2016useful}.

\section{Results}

\subsection{Sociodemographic parameters of the sample}

The \textit{Fadigômetro} questionnaire, containing sociodemographic, behaviour and health questions was filled up by 796 aircrew workers without distinction of sex, race, rank, age or years in the job from July 19, 2018 until March 02, 2021. Among all responders, 66.6\% (530) were male, 51.8\% pilots (412) and 48.2\% (384) flight attendants (cabin crew). Of the 412 pilots, 54.1\% (223) were captains and 45.9\% (189) first officers. The average ages and standard deviations (in years) were 40.8$\pm$9.5 (N=512), 35.6$\pm$7.1 (N=256), 41.8$\pm$9.9 (N=400), 36.0$\pm$7.0 (N=368), 47.1$\pm$9.2 (N=368) and 35.6$\pm$6.5 (N=184) for Male, Female, Pilots, Flight Attendants, Captains and First Officers, respectively. The number of validated responses for each group differs from the total number of responders since some of the responses were not provided.

\subsection{SAFTE-FAST outputs and \textit{productivity metrics}}

In this section we present all SAFTE-FAST outputs (software version 4.0.3.207), as well as our \textit{productivity metrics} for 30-day epochs rosters from January 2019 until February of 2020. A 15-day epoch was also used for mid-March of 2019 and 2020. 
Considering that several filters were adopted to extract the input CSV files (see Methods), the Ids whose rosters were analysed vary from month by month and represent a fraction of the eligible participants of the study. For this reason, the modelling results not necessarily include all the eligible Ids, neither all the Ids that contributed for the questionnaire, but the Ids that fulfilled the filter requirements defined previously.\\
The results of monthly averages and corresponding standard errors of the SF outputs for $EM_{C}$, $RM_{C}$, $FHA_{C}$ are presented in Table \ref{tab:table1}, together with our estimates of $SD^{max}$ and $t_{awake}^{max}$, calculated from Eqs. \ref{eqn:2} and \ref{eqn:3}, respectively. The first column of Table \ref{tab:table1} shows the total number of validated rosters for each period, which varies between 389 to 680 depending on the month, with a consistent increase since June of 2019. January 2019 presents the highest fatigue scores with an average $EM_{C}$ around 71.8\%, in contrast with June of 2019, which presented a much higher average of $\sim $ 78.1\%. Considering the average minimum sleep reservoir, January 2019 also presents the lowest score of $\sim $ 75.8\%, which is consistent with an average maximum sleep debt of almost 8 hours (7.75 h) and an average equivalent maximum time awake of almost 24 hours (23.25 h). Considering one standard error, the relative uncertainties of $EM_{C}$, $RM_{C}$, $FHA_{C}$ and $SD^{max}$ (or $t_{awake}^{max}$) vary typically from 0.30 to 0.51\%, 0.17 to 0.27\%, 5.4 to 9.3\% and 0.60 to 0.93\%, respectively; thus evidencing the high precision characteristic of our model estimates.

\begin{table*}
\caption{\label{tab:table1}
Average values and standard errors for the minimum effectiveness ($EM_{C}$), minimum sleep reservoir ($RM_{C}$), fatigue hazard area ($FHA_{C}$), maximum sleep deficit ($SD^{max}$) and maximum equivalent time wake ($t_{awake}^{max}$) during critical phases of flight from January 2019 up to February 2020 in 30-day epochs. Also shown the total number of rosters for each period (N) and the SF results for 15-day epochs (*) for mid-March of 2019 and 2020.}
\begin{ruledtabular}
\begin{tabular}{ccccccc}
Period &N &$\langle EM_{C}\rangle$ (\%) & $\langle RM_{C}\rangle$ (\%) & $\langle FHA_{C}\rangle$ (min) & $\langle SD^{max}\rangle$ (h) & $\langle t_{awake}^{max}\rangle$ (h)\\
\hline
Jan-19	&419	&71.75$\pm$0.28	&75.78$\pm$0.21	&8.01$\pm$0.48	&7.75$\pm$0.07	&23.25$\pm$0.20\\
Feb-19	&435	&74.13$\pm$0.33	&76.90$\pm$0.19	&5.50$\pm$0.36	&7.39$\pm$0.06	&22.17$\pm$0.19\\
Mar-19	&404	&75.72$\pm$0.39	&77.95$\pm$0.20	&4.73$\pm$0.38	&7.06$\pm$0.07	&21.17$\pm$0.20\\
Apr-19	&389	&75.73$\pm$0.35	&77.92$\pm$0.19	&4.01$\pm$0.32	&7.06$\pm$0.06	&21.19$\pm$0.18\\
May-19	&399	&76.57$\pm$0.34	&78.41$\pm$0.18	&3.30$\pm$0.31	&6.91$\pm$0.06	&20.73$\pm$0.17\\
Jun-19	&554	&78.11$\pm$0.35	&79.32$\pm$0.19	&2.85$\pm$0.21	&6.62$\pm$0.06	&19.86$\pm$0.18\\
Jul-19	&673	&74.04$\pm$0.26	&77.09$\pm$0.15	&6.02$\pm$0.35	&7.33$\pm$0.05	&21.99$\pm$0.15\\
Aug-19	&635	&74.92$\pm$0.25	&77.59$\pm$0.14	&4.37$\pm$0.27	&7.17$\pm$0.04	&21.52$\pm$0.13\\
Sep-19	&644	&74.92$\pm$0.28	&77.45$\pm$0.17	&4.79$\pm$0.30	&7.22$\pm$0.05	&21.65$\pm$0.16\\
Oct-19	&652	&75.01$\pm$0.28	&77.74$\pm$0.16	&5.04$\pm$0.33	&7.12$\pm$0.05	&21.37$\pm$0.15\\
Nov-19	&649	&76.04$\pm$0.29	&78.75$\pm$0.16	&3.92$\pm$0.25	&6.80$\pm$0.05	&20.40$\pm$0.15\\
Dec-19	&674	&74.58$\pm$0.25	&77.76$\pm$0.15	&4.79$\pm$0.30	&7.12$\pm$0.05	&21.35$\pm$0.15\\
Jan-20	&680	&73.94$\pm$0.22	&77.57$\pm$0.14	&5.29$\pm$0.28	&7.18$\pm$0.04	&21.53$\pm$0.13\\
Feb-20	&670	&75.29$\pm$0.26	&78.21$\pm$0.15	&4.32$\pm$0.26	&6.97$\pm$0.05	&20.92$\pm$0.14\\
Mar-19 (1/2)*& 	400	&78.45$\pm$0.40	&79.54$\pm$0.19	&2.55$\pm$0.24	&6.55$\pm$0.06	&19.64$\pm$0.19\\
Mar-20 (1/2)* 	&599	&80.09$\pm$0.34	&81.06$\pm$0.17	&1.69$\pm$0.15	&6.06$\pm$0.05	&18.18$\pm$0.16\\
\end{tabular}
\end{ruledtabular}
\end{table*}

Table \ref{tab:table2} presents the results of our \textit{productivity metrics} extracted by a dedicated filter algorithm for the same group of rosters that generated the SF fatigue outcomes of Table \ref{tab:table1}. As clearly shown in Table \ref{tab:table2}, these metrics present large variations from month to month, providing a workload profile for 2019 and early 2020. Once again, Jan-19 presents the highest scores for all metrics, except $N_{work}$, which is associated with non-crewing and mostly training activities. The typical high season months in South-America of Jan-19 and Jul-19 have the highest scores regarding the average number of night shifts ($6.56\pm0.11$ and $6.46\pm0.10$) and the average number of departures and landings between 2 and 6 a.m. ($4.55\pm0.15$ and $4.51\pm0.13$), respectively. On the other hand, Jan-20, which should also be considered a high season month, presents a lower score for $N_{NS}$ ($6.19\pm0.11$), when compared with Aug-19 ($6.27\pm0.10$) and Sep-19 ($6.23\pm0.11$). These results show that other factors beyond the trivial high/low seasonal variation do play a relevant role in the fatigue outcomes.\\

Shapiro-Wilk normality tests showed that the SF outputs $EM_{C}$, $RM_{C}$ and $FHA_{C}$, as well as \textit{the productivity metrics} $N_{NS}$, $DT$, $N_{crew}$ and $N_{wocl}$ are unlikely to be originated from a normal distribution ($\textit{p}<0.038$ in all cases). For this reason, non-parametric Mann-Whitney tests for independent samples were applied for the investigation of group effects when comparing the early months of 2019 and 2020. The results of the 2020/2019 ratios and the corresponding \textit{p}-values are depicted in Table \ref{tab:table3} for 30- (January and February) and 15- (March) day epochs for 2019 and 2020. Except for $N_{NS}$ and $N_{wocl}$ in the comparison between Feb-19\&Feb-20 (\textit{p} = 0.091 and 0.058, respectively) and Mar-19\&Mar-20 (\textit{p} = 0.279 and 0.159, respectively), all the other results show a quantitative decrease in the fatigue scores between early 2019 and 2020. The 2020/2019 ratios for $FHA_{C}$, $N_{NS}$, $DT$, $N_{crew}$ and $N_{wocl}$ vary between 0.66 to 0.79, 0.94 to 1.04, 0.89 to 0.94, 0.89 to 0.92 and 0.92 to 0.94, respectively.

\begin{table*}
\caption{\label{tab:table2}
Average values and standard errors for the number of night shifts ($N_{NS}$), number of consecutive night shifts ($N_{CNS}$), duty time ($DT$), number of \textit{Crewing} events ($N_{crew}$), number of \textit{Working} events ($N_{work}$) and number of departures and arrivals within 2 and 6 a.m. ($N_{wocl}$) from January 2019 until February 2020 in 30-day epochs. Also shown the same \textit{productivity metrics} for 15-day epochs (*) for mid-March of 2019 and 2020.}
\begin{ruledtabular}
\begin{tabular}{ccccccc}
Period & $\langle N_{NS}\rangle$ &$\langle N_{CNS}\rangle$ & $\langle DT\rangle (h)$ &$\langle N_{crew}\rangle$&$\langle N_{work}\rangle$ &$\langle N_{wocl}\rangle$\\
\hline
Jan-19&	6.56$\pm$0.11&	2.07$\pm$0.06&	116.77$\pm$1.17&	33.64$\pm$0.53&	2.73$\pm$0.12&	4.55$\pm$0.15\\
Feb-19&	5.87$\pm$0.13&	1.80$\pm$0.06&	110.60$\pm$1.54&	31.80$\pm$0.65&	2.98$\pm$0.12&	4.11$\pm$0.15\\
Mar-19&	5.26$\pm$0.14&	1.55$\pm$0.06&	105.65$\pm$1.56&	29.57$\pm$0.63&	3.68$\pm$0.15&	3.67$\pm$0.16\\
Apr-19&	5.68$\pm$0.13&	1.62$\pm$0.06&	110.88$\pm$1.48&	30.99$\pm$0.65&	3.92$\pm$0.16&	3.56$\pm$0.15\\
May-19&	5.36$\pm$0.13&	1.47$\pm$0.06&	108.12$\pm$1.54&	30.53$\pm$0.65&	4.41$\pm$0.15&	3.25$\pm$0.14\\
Jun-19&	4.25$\pm$0.13&	1.16$\pm$0.05&	82.51$\pm$2.05&	23.53$\pm$0.68&	3.23$\pm$0.13&	2.64$\pm$0.11\\
Jul-19&	6.46$\pm$0.10&	1.98$\pm$0.05&	113.13$\pm$1.01&	32.54$\pm$0.52&	4.63$\pm$0.16&	4.51$\pm$0.13\\
Aug-19&	6.27$\pm$0.10&	1.93$\pm$0.05&	113.93$\pm$1.07&	32.47$\pm$0.51&	5.37$\pm$0.17&	3.95$\pm$0.13\\
Sep-19&	6.23$\pm$0.11&	1.98$\pm$0.05&	107.13$\pm$1.23&	30.44$\pm$0.52&	5.28$\pm$0.17&	4.08$\pm$0.13\\
Oct-19&	6.06$\pm$0.11&	1.85$\pm$0.05&	107.33$\pm$1.21&	30.83$\pm$0.53&	5.36$\pm$0.16&	3.99$\pm$0.13\\
Nov-19&	5.24$\pm$0.11&	1.55$\pm$0.05&	101.80$\pm$1.38&	27.79$\pm$0.54&	5.49$\pm$0.19&	3.21$\pm$0.11\\
Dec-19&	5.81$\pm$0.10&	1.83$\pm$0.06&	108.21$\pm$0.98&	29.80$\pm$0.46&	5.19$\pm$0.17&	3.61$\pm$0.13\\
Jan-20&	6.19$\pm$0.11&	2.01$\pm$0.06&	109.98$\pm$1.01&	30.99$\pm$0.49&	4.60$\pm$0.16&	4.28$\pm$0.13\\
Feb-20&	5.62$\pm$0.10&	1.68$\pm$0.05&	103.68$\pm$1.12&	28.30$\pm$0.48&	5.22$\pm$0.19&	3.79$\pm$0.12\\
Mar-19(1/2)*&  2.81$\pm$0.09&	0.78$\pm$0.04& 56.19$\pm$0.80&	15.64$\pm$0.33&1.91$\pm$0.09&	2.03$\pm$0.11\\
Mar-20(1/2)*& 	2.91$\pm$0.07&	0.85$\pm$0.03&	50.23$\pm$0.80&	14.43$\pm$0.32&	2.57$\pm$0.13&	1.88$\pm$0.10\\  
\end{tabular}
\end{ruledtabular}
\end{table*}

\begin{table*}
\caption{\label{tab:table3}
2020/2019 ratios and Mann-Whitney tests for independent samples (\textit{p}-values) for $EM_{C}$, $RM_{C}$, $FHA_{C}$, $N_{NS}$, $DT$, $N_{crew}$ and $N_{wocl}$ in 30-day epochs for early 2019 and 2020. Also shown the comparison of 15-day epochs (*) for March 2019 and 2020.}
\begin{ruledtabular}
\begin{tabular}{ccccc}

\multirow{2}{*}{Variables} & \multirow{2}{*}{Parameters}& \multicolumn{3}{c}{Groups}\\
& & Jan-19\&Jan-20 & Feb-19\&Feb-20 & Mar-19 (1/2)\&Mar-20 (1/2)*\\
\hline

\multirow{2}{*} {$EM_{C}$}& 2020/2019 ratio & 0.970	&0.985	&0.980\\
 & \textit{p}-value & $<0.001$& 0.001 & 0.004\\
\multirow{2}{*}{$RM_{C}$}& 2020/2019 ratio &0.977	&0.983	&0.981\\
&\textit{p}-value & $<0.001$& $<0.001$& $<0.001$\\	
\multirow{2}{*}{$FHA_{C}$} & 2020/2019 ratio & 0.66 &	0.79&0.66\\
&\textit{p}-value &$<0.001$&0.004&$<0.001$\\
\multirow{2}{*}{$N_{NS}$} & 2020/2019 ratio &0.94	&0.96&1.04\\	
&\textit{p}-value &0.037	&0.091	&0.279\\	
\multirow{2}{*}{$DT$} & 2020/2019 ratio &0.94&	0.94&0.89\\
& \textit{p}-value & $<0.001$&$<0.001$&$<0.001$\\
\multirow{2}{*}{$N_{crew}$} & 2020/2019 ratio &0.92	&0.89&0.92\\
&\textit{p}-value &0.001&	$<0.001$&0.010	 \\
\multirow{2}{*}{$N_{wocl}$} & 2020/2019 ratio&0.94	&0.92&0.93\\
&\textit{p}-value &0.041&0.058&0.159\\
\end{tabular}
\end{ruledtabular}
\end{table*}
  
\subsection{Potential root causes of fatigue}  

\subsubsection{Average minimum SF Effectiveness versus $N_{NS}$} 
The minimum SF effectiveness during critical phases of flight ($EM_{C}$) represents the worst fatigue score (hot spot) for a given subject during a given period of analysis. For this reason, it is likely that crew members with the same amount of night shifts would have similar values of $EM_{C}$, since the latter depends on the sleep opportunities of the rosters, which are adversely affected by night shifts. In that sense, crew members with a large number of night shifts are likely to have lower minimum effectiveness scores in critical phases of flight.\\
The relationship between the average $EM_{C}$, herein denoted as $<EM_{C}>$, and $N_{NS}$ is shown in the upper panel of Fig.\ref{fig:Fig1}, which includes all 2019 rosters for 30-day epochs with $1\leq N_{NS}\leq 13$. The average values of $<EM_{C}>$ (data points) are satisfactorily fitted by a third degree polynomial, represented by the solid blue line ($\chi^2$ = 9.00,\textit{ d.o.f.} = 9 and \textit{p} = 0.437). The dashed-dotted red and green lines represent, respectively, the upper and lower limits considering a 95\% confidence interval (CI). The fitting was performed using the least square method \citep{helene2016useful} and the confidence intervals were obtained using uncertainty propagation techniques and the covariance matrix of the best-fit parameters, as similarly described elsewhere \citep{rodrigues2020monte}. It is verified that $<EM_{C}>$ drops more significantly for $1<N_{NS}\lesssim 5$ and $N_{NS}\gtrsim 10$. The error bars are higher for $N_{NS}\geq 10$, since there are fewer rosters within this range.
\subsubsection{Relative Fatigue Risk versus $N_{NS}$}
\begin{center}
\begin{figure}
\includegraphics[scale=0.45]{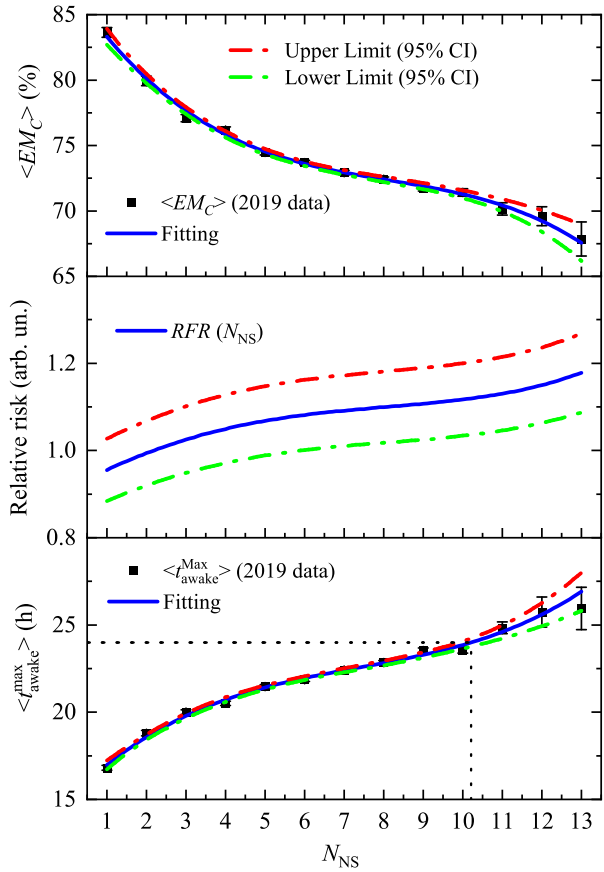}
\caption{\label{fig:Fig1}Upper Panel: Average values and standard errors of $EM_{C}$ (data points) as a function of $N_{NS}$ and its corresponding third degree polynomial fitting (solid blue line). Center panel: Relative Fatigue Risk (\textit{RFR}) as a function of $N_{NS}$ (blue line). Lower panel: Average equivalent maximum time awake during critical phases of flight as a function of $N_{NS}$ (data points), its corresponding third degree polynomial fitting (solid blue line) and a 24 hour time awake reference (dotted black line). In all panels the dashed-dotted red and green lines represent the upper and lower limits, respectively, considering a 95\% CI.}
\end{figure}
\end{center}

Considering that the probability of human factor accidents vary inversely with $E_{SF}$ (see Eq.\ref{eqn:1}), one can estimate the relative fatigue risk as a function of $N_{NS}$ by $RFR(x)\cong b/f(x)$, where $f(x)$ stands for the fitted function and $x\equiv N_{NS} $. Under this approximation, one can also calculate the 95\% CI's of $RFR(x)$ by the propagation of the uncertainties of $f(x)$ and \textit{b}. The results of this relative risk estimate and its upper/lower limits are shown by the solid blue and dashed-dotted red/green lines of the center panel of Fig.\ref{fig:Fig1}, respectively. As expected, the relative fatigue risk increases with $N_{NS}$. Increasing the number of night shifts from 1 to 13 increases the relative risk by 23.3\% (95\% CI, 20.4 - 26.2\%).

\subsubsection{Average maximum equivalent time awake versus $N_{NS}$}  
The lower panel of Fig.\ref{fig:Fig1} presents the average maximum equivalent time awake ($t_{awake}^{max}$) - associated with the average minimum SF sleep reservoir during critical phases of flight - as a function of $N_{NS}$ (data points) and its corresponding standard errors (error bars). Once again, a third degree polynomial fitting ($\chi^2$ = 16.32,\textit{ d.o.f.} = 9 and \textit{p} = 0.060) successfully describes the data. It is also verified that $t_{awake}^{max}$ has a higher slope for $N_{NS}\geq10$, exceeding more than 24 hours of equivalent wakefulness for $N_{NS}\geq11$.

 \begin{table*}
\caption{\label{tab:table4}
Best-fit parameters and fitting results for $\langle EM_{C}\rangle $, $\langle t_{awake}^{max}\rangle$ and $\langle FHA_{C}\rangle$.}
\begin{ruledtabular}
\begin{tabular}{ccccc}

\multicolumn{2}{c}{Fitting} & $\langle EM_{C}\rangle$&$\langle t_{awake}^{max}\rangle$& $\langle FHA_{C}\rangle$\\
\hline
& \multirow{2}{*}{Model} & \multicolumn{2}{c}{$f(x) = a + bx + cx^{2}+dx^{3}$} &$f(x) = a + bx + cx^{2}$\\
& & \multicolumn{2}{c}{$x\equiv N_{NS}$} & $x\equiv N_{wocl}$\\
\hline
\multirow{7}{*}{\rotatebox[origin=c]{90}{Results}} & \textit{a} &87.4 $\pm$ 0.5 \%	&	14.95 $\pm$ 0.25 h	&	0.246 $\pm$ 0.028 min\\
&\textit{b}& -4.55 $\pm$ 0.31 \%	&	2.26 $\pm$ 0.17 h	&	0.468 $\pm$ 0.035 min\\	
&\textit{c}& 0.50 $\pm$ 0.05 \%	&	-0.250 $\pm$ 0.032 h	&	0.115 $\pm$ 0.005 min\\
&\textit{d}&	-0.0204 $\pm$ 0.0028 \%	&	0.0113 $\pm$ 0.0018 h	&	--\\
& $\chi^2$ & 9.00	&	16.32	&	16.68\\
& \textit{d.o.f.} & 	9	&	9	&	14\\
& \textit{p}-value	& 0.437	&	0.060	&	0.274\\	

 \end{tabular}
\end{ruledtabular}
\end{table*}
 
\subsubsection{Fatigue hazard area versus $N_{wocl}$}  
\begin{center}
\begin{figure}
\includegraphics[scale=0.37]{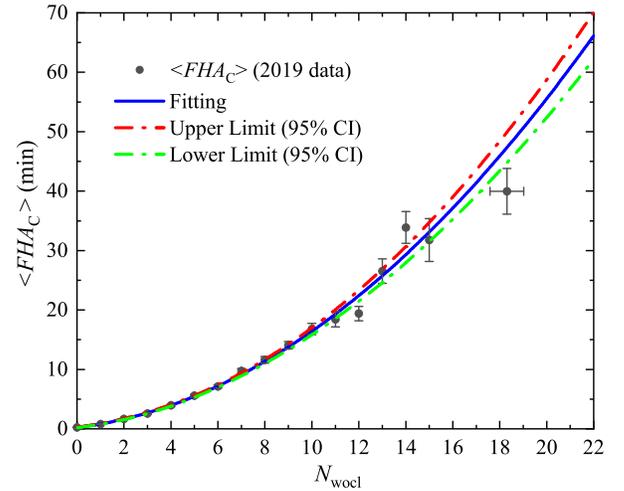}
\caption{\label{fig:Fig2}Average fatigue hazard area in critical phases of flight as a function of $N_{wocl}$ (data points). The solid blue line represents a second degree polynomial fitting, where the dashed-dotted red and green lines represent, respectively, the upper and lower limits, considering a 95\% CI.}
\end{figure}
\end{center}

The cumulative $FHA_{C}$ is expected to increase with the number of departures and landings within the WOCL period, representing a consistent overall fatigue score for a given subject in a given period of analysis. Fig.\ref{fig:Fig2} shows the average values of $FHA_{C}$ as a function of $N_{wocl}$ (data points) and their respective standard errors (error bars), for all the 6527 rosters of 2019. At this time, the data are consistently fitted ($\chi^2$ = 16.68,\textit{ d.o.f.} = 14 and \textit{p} = 0.274) by a parabolic function (solid blue line), with the dashed-dotted red/green lines representing the upper/lower limits, respectively, considering a 95\% CI. The data point at $x = 18.3\pm0.7$ and $y = 40.0\pm3.9$ min represents \textit{x}- and \textit{y}-average values for all rosters with $16 \leq N_{wocl}\leq 22$. The x-axis error (0.7) corresponds to the standard error of the \textit{x}-average and was propagated to the y-axis error (3.9 min) in order to perform the fitting procedure. All the fitting results shown in Figs. \ref{fig:Fig1} and \ref{fig:Fig2} are summarized in Table \ref{tab:table4}.
   
\subsection{Modelling the monthly-averaged fatigue hazard area} 
\begin{center}
\begin{figure*}
\includegraphics[scale=0.65]{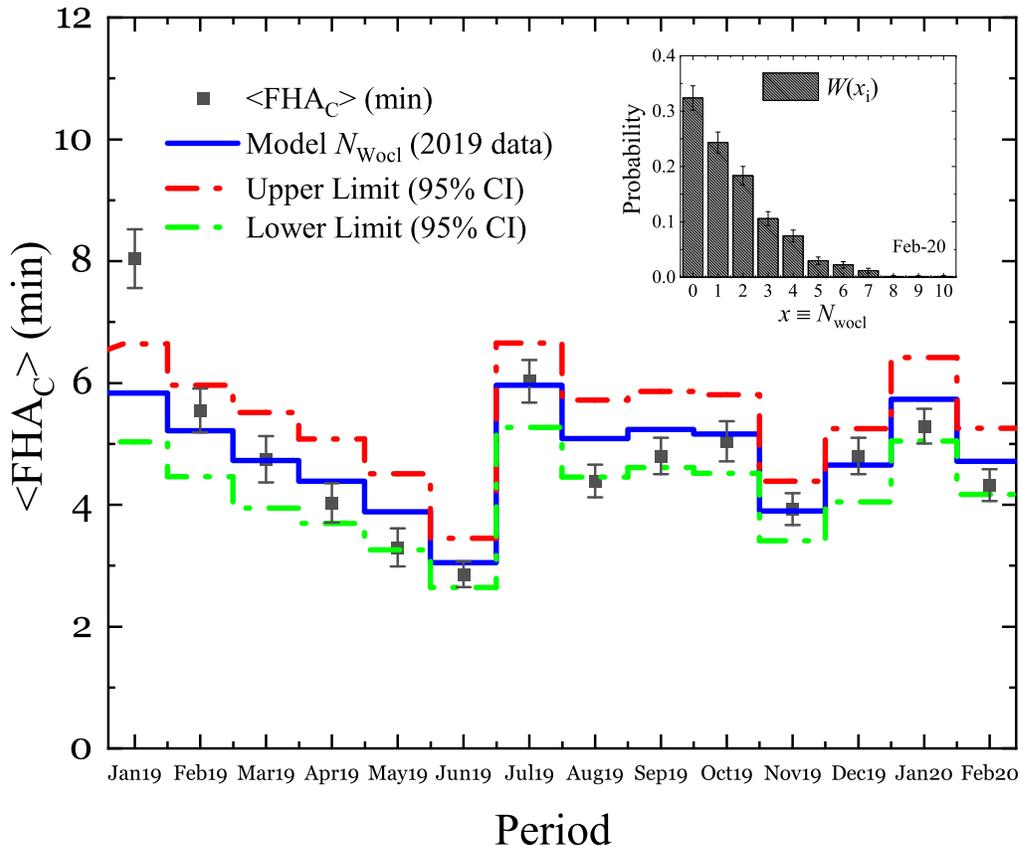}
\caption{\label{fig:Fig3} $\langle FHA_{C} \rangle$ and its corresponding standard errors for each 30-day epochs (data points) and the model predictions based on the $N_{wocl}$ distributions of each period (solid blue line). The dashed-dotted red and green lines represent the upper and lower limits, respectively, considering a 95\% CI.}
\end{figure*}
\end{center}

The average $FHA_{C} $ in 30-day epochs presented in Table \ref{tab:table1} represents an overall fatigue metric for any given set of rosters. In this regard, a suitable model to estimate $\langle FHA_{C}\rangle$ for a given $N_{wocl}$ distribution is highly desirable, given its practical relevance to guide airline policies and management strategies for those involved in crew rostering processes.
Consequently, the monthly-averaged fatigue hazard area in critical phases of flight can be written as:
\begin{equation}
\label{eqn:5}
\langle FHA_{C}^{j} \rangle= \sum_{i=0}^{i_{max}} W^{j}(x_{i})f(x_{i}),
\end{equation}
where $W^{j}(x_{i})$ represents the normalized $N_{wocl}$ distribution for a given month \textit{j}, $f(x_{i})$ the fitted parabola of Fig.\ref{fig:Fig2}, both calculated at each $x_{i}$ value, with the sum going from zero up to $x_{i_{max}}$, which represents the maximum $N_{wocl}$ of the distribution. Fig. \ref{fig:Fig3} presents our model estimates for $\langle FHA_{C} \rangle$ (blue line) with its upper and lower 95\% CI given by the dashed-dotted red and green lines, respectively. The insert of Fig. \ref{fig:Fig3} shows, as an example, the normalized $N_{wocl}$ distribution for all the 670 rosters of Feb-2020. The 95\% CI of $\langle FHA_{C}^{j} \rangle$ was calculated propagating the uncertainties of $f(x_{i})$ and $W^{j}(x_{i})$ at each $x_{i}$ value. The latter is assumed as $\frac{\sqrt{n^{j,i}}}{n^{j,i}}W^{j}(x_{i})$, with $n^{j,i}$ representing the number of events for each $x_{i}$ value at a given \textit{j} month. As observed in Fig. \ref{fig:Fig3}, the model calculations do reproduce the several structures that appear in $\langle FHA_{C} \rangle$, with all data falling within the 95\% CI, except for Jan-19, which is considerably higher than the model predictions. This latter can be explained considering that January 2019 has a $\langle FHA_{C} \rangle$ score substantially higher than the average figures of 2019, which were used as a baseline do determine the polynomial best-fit parameters. Both January and February of 2020 are also well reproduced by our model estimates, despite of not being included in the fitting shown in Fig. \ref{fig:Fig2}, to avoid any seasonal bias in mixing two months of 2020 with our annual based metric for 2019.

\subsection{Circadian variations of fatigue outcomes and associated risk}          

\begin{center}
\begin{figure}
\includegraphics[scale=0.36]{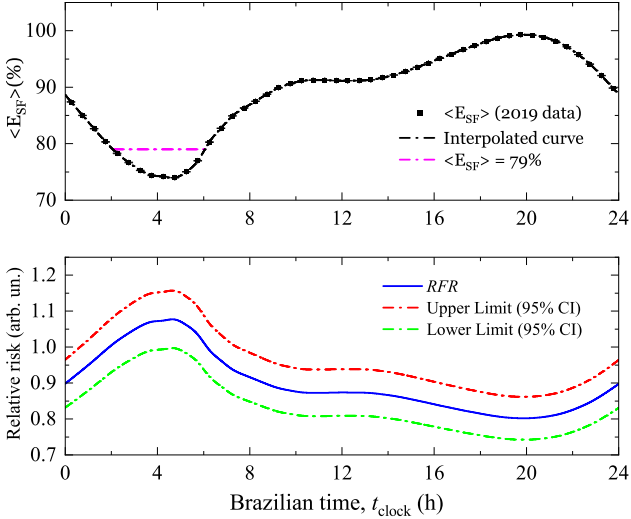}
\caption{\label{fig:Fig4}Upper panel: Average SF effectiveness scores (black squares) and its standard errors (error bars) for all the 30-min assessments during all crewing events of 2019. The dashed-dotted black line is only to guide the eyes and the dashed-dotted magenta line corresponds to $\langle E_{SF}\rangle = 79\%$. Lower panel: Model estimates for the \textit{RFR} as a function of the time of the day (solid blue line) and its upper (dashed-dotted red) and lower (dashed-dotted green) limits considering a 95\% CI.}
\end{figure}
\end{center}

\begin{center}
\begin{figure}
\includegraphics[scale=0.36]{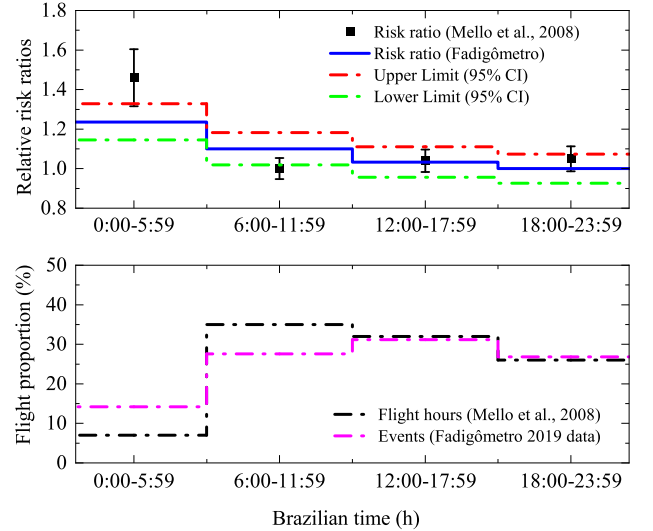}
\caption{\label{fig:Fig5}Upper panel: Relative risk ratios from Mello et al. 2008 (data points) in comparison with the \textit{Fadigômetro} predictions (solid blue histogram) and its upper (dashed-dotted red) and lower (dashed-dotted green) limits considering a 95\% CI. Lower panel: Flight proportion (\%) reported by Mello et al. 2008 (dashed-dotted black histogram) versus the fraction of events from the Fadigômetro data (dashed-dotted magenta histogram).}
\end{figure}
\end{center}

The SAFTE-FAST software provides effectiveness scores for all the individuals at each 30 minute time interval, allowing the verification of circadian oscillations in $E_{SF}$ as a function of the time of the day. The results of this analysis - which includes all the 742521 30-min crewing events of 2019 - are presented in the upper panel of Fig. \ref{fig:Fig4}. The solid black squares represent the average values of $E_{SF}$ for each 30 min bin, considering all effectiveness scores of all individuals during all flights of 2019. The error bars represent the standard errors and range from 0.013 up to 0.056\%, showing the high precision of $\langle E_{SF}\rangle$. The dashed-dotted black line is an interpolated curve only to guide the eyes, whereas the dashed-dotted magenta line corresponds to $\langle E_{SF}\rangle = 79\%$, which occurs at $02h04 $ and $06h03$, considering the Brazilian legal time. Such time interval encompasses the worst fatigue outcomes for all the 30-min crewing assessments, reinforcing our choice to establish the WOCL events within 02h00 and 06h00. As verified, the average effectiveness varies quite significantly as a function of $t_{clock}$ and drops below 90\% between 11:45 p.m. and 9:15 a.m.
The lower panel of Fig. \ref{fig:Fig4} shows our model estimate for the relative fatigue risk as a function of $t_{clock}$ (solid blue line) calculated as $RFR(t_{clock})\sim P_{HF}(t_{clock}) \cong b/\langle E_{SF}(t_{clock})\rangle $. The dashed-dotted red and green lines, represent, respectively, the upper and lower limits considering a 95\% CI and were obtained propagating the uncertainty of \textit{b}. The uncertainty of $\langle E_{SF}\rangle$ in the \textit{RFR} was not taken into account, given its negligible values ($\leq0.056\%$).\\
In order to compare our model predictions for the relative fatigue risk ratios as a function of the time of the day with previous measurements of pilot errors in the cockpit \citep{Mello2008} we firstly averaged the continuous function $RFR(t_{clock})$ within the same time intervals investigated in Ref. \citep{Mello2008}. Secondly, we normalized our results by equalling our lowest \textit{RFR} average (within 18:00 and 23:59) to unit. The results of these procedures are presented by the solid blue histogram in the upper panel of Fig. \ref{fig:Fig5}, together with the upper (dashed-dotted red) and lower (dashed-dotted green) limits considering a 95\% CI and the normalized risk ratios found by Mello et al. 2008 \citep{Mello2008} (data points). For the latter, we have also assumed an error bar proportional to the ratio of $ \frac{\sqrt{N}}{N}$, where \textit{N} stands for the absolute number of errors within a given time interval. Such approximation holds if the probability of errors follow a Poisson distribution. As verified, our estimates agree qualitatively with the objective measurements performed in Ref. \citep{Mello2008} (see Limitations Section). The lower panel of Fig. \ref{fig:Fig5} shows the proportion of flights as a function of the time of the day reported by Mello et al., 2008 \citep{Mello2008} (dashed-dotted black histogram) in comparison with our estimates for the proportion of events using all 2019 data (dashed-dotted magenta histogram). As shown in the lower panel of Fig. \ref{fig:Fig5}, the risk exposure presents a significant increase (from 7 to 14\%) within 0h00 and 5:59, comparing the flight proportion found in 2005 (when the data of Ref. \citep{Mello2008} were collected) with 2019.

\subsection{Tailored analyses of SAFTE-FAST inputs}

In this section we demonstrate quantitative variations in some SAFTE-FAST outputs when tailoring some key input metrics related to afternoon naps, commuting, and bedtime constraints.\\
For the afternoon naps, the SAFTE-FAST console has a standard input parametrization (Auto-Nap) that assumes a nap before a night duty, which depends on the hours of sustained wakefulness since the last sleep event (see Supplementary Section). This particular input can be switched off, which means that no afternoon nap will be added automatically, regardless of the time period since the last sleep event (Auto-Nap OFF).\\  
For the commuting from home to base station and vice-versa, we have adopted a \textit{standard} metric of one hour and also an \textit{extended} commuting profile of 2 hours. So, for check-in purposes, we consider one hour of preparation at home, hotel or rest facility and one (\textit{standard}) or two (\textit{extended}) hours of commuting from home to station and vice-versa.\\ For the bedtime parameter, the SAFTE-FAST algorithm adopts 11 p.m. as default, which is also adopted in our calculations. However, the model also has an advance bedtime function, i.e. a feature that assumes an individual will go to bed earlier than usual should an early start shift is scheduled in the following morning, regardless of the sleep deficit accumulated in previous shifts. This feature can also be switched off, which means that the standard bedtime of 11 p.m. will be a fixed constraint for all main sleep events during night.\\
The quantitative effects in few SAFTE-FAST outputs related with the variations of these input criteria are shown in Table \ref{tab:table5}, considering two months with low (May and June of 2019) and high (February and July of 2019) productivity profiles. The calculations were done using the fractions of the monthly rosters related with the Ids who declared that are not used to get afternoon naps prior to night shifts (within 42.2 to 45.4\% of the responders), two hours of commuting from home to station and vice-versa (22.6 to 25.3\% of the responders), or not used to advance the bedtime before early-starts (32.1 to 33.3\% of the responders). Wilcoxon signed-rank tests for matched pairs show significant group effects ($ p<0.001$) for $EM_{C}$, $RM_{C}$ and $FHA_{C}$ comparing the groups with and without the afternoon nap, with one or two hours of commuting and with or without the advanced bedtime feature. Pearson's correlations (effect sizes), with and without the the Auto-Nap function enabled, are equal or higher than 0.967 (0.765), 0.926 (0.674), and 0.941 (0.544) for $EM_{C}$, $RM_{C}$ and $FHA_{C}$, respectively. For the comparison between one and two hours of commuting, Pearson's correlations (effect sizes) are equal or higher than 0.970 (0.409), 0.912 (0.530) and 0.921 (0.300) for $EM_{C}$, $RM_{C}$ and $FHA_{C}$, respectively. For the groups with and without the advanced bedtime feature, Pearson's correlations (effect sizes) are equal or higher than 0.970 (0.714), 0.926 (0.852) and 0.894 (0.418) for $EM_{C}$, $RM_{C}$ and $FHA_{C}$, respectively.  The average fatigue hazard area in critical phases of flight increases by 43 to 63\% when switching off the afternoon naps, 14 to 21\% when increasing the commuting time from home to station and vice-versa from one to two hours and 35 to 54\% when switching off the advanced bedtime criterion.      
  
\begin{table*}
\caption{\label{tab:table5} Average values and standard deviations (SD) for $EM_{C}$, $RM_{C}$ and $FHA_{C}$ with or without the Auto-Nap function, with one or two hours of commuting and with or without the Advanced bedtime (BT) feature for February, May, June and July of 2019. Also shown the total number of responders (N) for each period and input, the corresponding fraction of responders (\%), as well as the \textit{p}-values, the Pearson's correlations, the effect sizes (dz) and the ratios of $\langle FHA_{C}\rangle$ for each matched pair of the analyses.}
\begin{ruledtabular}
\begin{tabular}{ccccccccccccccc}

\multirow{2}{*}{SF Input} & &  & \multicolumn{4}{c}{$EM_{C} (\%)$} &\multicolumn{4}{c}{$ RM_{C} (\%)$} & \multicolumn{4}{c}{$FHA_{C}(min)$} \\
& & & Feb-19 & May-19& Jun-19 & Jul-19& Feb-19 & May-19& Jun-19 & Jul-19& Feb-19 & May-19& Jun-19 & Jul-19\\
\hline
\multirow{10}{*}{Auto-Nap} &\multirow{2}{*}{ON} & Average & 73.76 & 76.54 &78.29	&74.07	&76.92	&78.35	&79.44	&77.19	&5.52	&3.27	&2.67	&5.64\\
& & SD &7.33	&6.60	&8.28	&6.49	&4.22	&3.52	&4.48	&3.70	&6.50	&6.68	&4.69	&6.85\\
&\multirow{2}{*}{OFF} & Average &72.15	&75.22	&76.74	&72.16	&75.80	&77.45	&78.32	&75.83	&8.20	&4.69	&4.35	&8.95\\
& & SD &8.07	&7.24	&9.24	&7.35	&4.58	&3.65	&5.02	&4.10	&8.86	&7.91	&7.11	&9.73\\
&  \multicolumn{2}{c}{N} & 194	&181	&234	&288	&194	&181	&234	&288	&194	&181	&234	&288\\
&  \multicolumn{2}{c}{Fraction (\%)} &44.6	&45.4	&42.2	&42.8	&44.6	&45.4	&42.2	&42.8	&44.6	&45.4	&42.2	&42.8\\
&  \multicolumn{2}{c}{\textit{p}-value}\footnotemark[1] & \multicolumn{4}{c}{$<0.001$} & \multicolumn{4}{c}{$<0.001$} & \multicolumn{4}{c}{$<0.001$}\\
&  \multicolumn{2}{c}{Pearson's $\rho $} & 0.967	&0.976	&0.986	&0.976	&0.940	&0.931	&0.956	&0.926	&0.960	&0.950	&0.957	&0.941\\
&  \multicolumn{2}{c}{Effect size, dz}\footnotemark[2] &0.763	&0.803	&0.886	&1.097	&0.716	&0.673	&0.743	&0.877	&0.840	&0.545	&0.569	&0.823\\
& & & \multicolumn{8}{c}{Ratio OFF/ON} &1.48	&1.43&	1.63&	1.59\\
\hline
\multirow{10}{*}{Commuting} &\multirow{2}{*}{1 hour} & Average & 73.90	&76.48	&78.13	&74.30	&77.11	&78.21	&79.38	&77.52	&4.57	&4.07	&2.66	&5.35\\
& & SD & 6.20	&7.63	&8.05	&6.90	&3.70	&3.96	&4.15	&3.83	&4.93	&6.22	&4.70	&6.60\\
&\multirow{2}{*}{2 hours} & Average &73.11	&75.62	&77.21	&73.51	&76.17	&77.33	&78.44	&76.58	&5.51	&4.64	&3.04	&6.36\\
& & SD & 5.95	&7.24	&7.84	&6.72	&3.65	&3.96	&4.32	&3.93	&5.75	&7.16	&5.15	&7.61\\
&  \multicolumn{2}{c}{N} & 99	&101	&132	&152	&99	&101	&132	&152	&99	&101	&132	&152\\
&  \multicolumn{2}{c}{Fraction (\%)} &22.8	&25.3	&23.8	&22.6	&22.8	&25.3	&23.8	&22.6	&22.8	&25.3	&23.8	&22.6\\
&  \multicolumn{2}{c}{\textit{p}-value}\footnotemark[1] & \multicolumn{4}{c}{$<0.001$} & \multicolumn{4}{c}{$<0.001$} & \multicolumn{4}{c}{$<0.001$}\\
&  \multicolumn{2}{c}{Pearson's $\rho $} & 0.973&	0.980	&0.970	&0.973	&0.929	&0.941	&0.912	&0.921	&0.921	&0.983	&0.972	&0.959\\
&  \multicolumn{2}{c}{Effect size, dz}\footnotemark[2] &0.551	&0.560	&0.470	&0.496	&0.678	&0.647	&0.527	&0.608	&0.414	&0.368	&0.304	&0.446\\
& & & \multicolumn{8}{c}{Ratio 2 hours/1 hour} &1.21	&1.14	&1.14	&1.19\\
\hline
\multirow{10}{*}{Advanced BT} &\multirow{2}{*}{ON} & Average &73.50	&76.62	&78.57	&74.36	&76.43	&78.17	&79.35	&77.20	&6.45	&3.68	&2.81	&6.18\\
& & SD &6.90	&6.95	&7.99	&7.22	&4.19	&3.46	&4.39	&4.18	&9.08	&7.61	&5.31	&9.65\\
&\multirow{2}{*}{OFF} & Average &72.04	&75.24	&77.18	&73.06	&74.88	&76.84	&78.00	&75.85	&9.50	&4.96	&4.31	&8.37\\
& & SD & 6.76	&6.81	&8.11	&7.09	&4.55	&3.82	&4.83	&4.34	&12.08	&4.96	&7.12	&12.76\\
&  \multicolumn{2}{c}{N} &145	&128	&178	&217	&145	&128	&178	&217	&145	&128	&178	&217\\
&  \multicolumn{2}{c}{Fraction (\%)} &33.3	&32.1	&32.1	&32.2	&33.3	&32.1	&32.1	&32.2	&33.3	&32.1	&32.1	&32.2\\
&  \multicolumn{2}{c}{\textit{p}-value}\footnotemark[1] & \multicolumn{4}{c}{$<0.001$} & \multicolumn{4}{c}{$<0.001$} & \multicolumn{4}{c}{$<0.001$}\\
&  \multicolumn{2}{c}{Pearson's $\rho $} &0.971	&0.970	&0.971	&0.974	&0.933	&0.926	&0.945	&0.942	&0.894	&0.970	&0.890	&0.930\\
&  \multicolumn{2}{c}{Effect size, dz}\footnotemark[2] &0.884	&0.816	&0.716	&0.794	&0.946	&0.921	&0.849	&0.925	&0.537	&0.551	&0.441	&0.422\\
& & & \multicolumn{8}{c}{Ratio OFF/ON} &1.47	&1.35	&1.54	&1.35\\

 \end{tabular}
\end{ruledtabular}
\footnotetext[1]{Wilcoxon signed-rank tests for matched pairs performed with SPSS version 25}
\footnotetext[2]{Effect size dz calculated with G*Power version 3.1.9.7 \citep{faul2007g}}
\end{table*}

\section{Discussion}
The SAFTE-FAST model outputs and most of the \textit{productivity metrics} showed a consistent reduction in the comparison between early 2019 and 2020, as depicted in Table \ref{tab:table3}. However, such finding does not allow an unambiguous conclusion about possible improvements of fatigue management policies from the operators, given that the \textit{productivity metrics} of $DT$ and $N_{crew}$ are also lower ($ p<0.010$) for early 2020. So, it is likely that this workload reduction reflected positively in the model outputs of $EM_{C}$, $RM_{C}$ and $FHA_{C}$ ($ p<0.001$ in all cases), but did not change quite substantially the root causes of fatigue given by $N_{NS}$ and $N_{wocl}$ for February ($ p=0.091$ and 0.058) and mid-March ($ p=0.279$ and 0.159), respectively. These results show that the SF outputs are quite sensitive to minor changes in $N_{NS}$ and $N_{wocl}$. For instance, for January 2020, the average value of $N_{wocl}$ is 6\% lower than for 2019, but the average  $FHA_{C}$ is 34\% lower. For mid-March 2020, the average $N_{wocl}$ is 7\% lower than the 2019 average, but without a significant group effect ($ p=0.159$). On the other hand, the average $FHA_{C}$ dropped 34\% in the same pair of samples ($ p<0.001$).\\    
The fatigue indicators of $EM_{C}$ and $t_{awake}^{max}$ show non-linear relationships with the \textit{productivity metric} of the number of night shifts $N_{NS}$ (see Table \ref{tab:table4}), which has an adverse effect on the flight schedules and drives the worst fatigue scores. Exceeding 10 night shifts in a 30-day time interval causes an average maximum equivalent time awake higher than 24 hours, or, equivalently, an average maximum sleep deficit of more than 8 hours. Indeed, Lamond and Dawson have found \citep{lamond1999} that 20-25 hours of continuous wakefulness can be associated with decrements in performance scores related with reasoning and attention, and that these degradations could be comparable for individuals with a blood alcohol concentration of 0.10 \%. The adverse effects of long periods of wakefulness were also pointed as a probable  cause for aviation accidents. The National Transportation Safety Board (NTSB) concluded that fatigue was a probable cause for the accident of American International Airways flight 808 in Guantanamo Bay on August 18, 1993 \citep{NTSB1993}. The NTSB final report determined that "...the probable causes of this accident were the impaired judgement, decision-making, and flying abilities of the Captain and flightcrew due to the effects of fatigue...". In fact, the analyses of the sleep/wake periods revealed that the Captain had been awake for 23.5 hours within the 28.5 hours prior to the accident \citep{NTSB1993, rosekind1993}. Consequently, our finding supports the recommendation that flight schedules should be planned with the lowest achievable number of night shifts, and not exceeding $N_{NS}=10$ within a 30-day time interval.\\
The $FHA_{C}$ also presents a non-linear relationship with the number of departures and landings within 2 and 6 a.m., $N_{wocl}$ (see Table \ref{tab:table4}). This result shows that cumulative fatigue builds up quadratically with the number of WOCL operations, reinforcing its adverse impact on the overall fatigue score for a given individual within a given time interval. Our model estimate for the monthly averaged $FHA_{C}$ for all the 30-day epochs of 2019 reproduces quite reasonably, except for January 2019, several structures that appear in the data (see Fig. \ref{fig:Fig3}), which present huge variations month by month. Consequently, our model approach allows the calculation of $\langle FHA_{C}\rangle$ for any given $N_{wocl}$ distribution, representing a suitable method for fatigue risk assessment. For the specific case of January 2019, however, our model is not able to reproduce the datum, which is significantly above the average figures found in 2019.\\
The $N_{wocl}$ distributions (see the insert of Fig. \ref{fig:Fig3} for a typical distribution of February 2020) drive the cumulative fatigue in rosters and should be concentrated within the lowest possible $N_{wocl}$ values. In fact, among all 2019 rosters, only 0.4\% present $16\leq N_{wocl}\leq 22$ and are associated with an average $FHA_{C}$ of $40.0 \pm 3.9$ min (see Fig. \ref{fig:Fig2}). Consequently, it is highly recommended that scores of $N_{wocl}>15$ are avoided in rosters, given its huge impact on overall fatigue and its negligible effect on crew productivity. Furthermore, only 3.6\% of the rosters had more than 10 WOCL operations within a 30-day time interval, giving room for safety improvements with minimal operational impact. Both the number of night shifts and the number of WOCL operations deserve a special attention when building crew rosters and operators should be encouraged to adopt them as key performance indicators to drive management policies and, whenever possible, rostering optimization processes. In order to avoid cumulative fatigue, these optimization processes should consider a dynamical evaluation of continuous 30-day time periods. \\
The relative fatigue risk as a function of the number of night shifts represents the inverse function of $EM_{C}$ and $N_{NS}$, times the parameter \textit{b} (see the central panel of Fig. \ref{fig:Fig1}). Increasing $N_{NS}$ from 1 to 10 increases the relative risk by 16.9\% (95\% CI, 16.0-17.8\%). Additionally, increasing $N_{NS}$ from 10 to 13 increases the risk by 5.5\% (95\% CI, 3.5-7.5\%). Once again, it is quite evident the safety benefit of avoiding more than 10 night shifts within a 30 days time interval.\\
The SF effectiveness as a function of the time of the day shows a relevant decrease within 11:45 p.m and 09:15 a.m, reaching its maximum and minimum scores around 8 p.m. and 4 a.m., respectively. Moreover, the SF effectiveness drops below 79\% within the WOCL period (from 2 to 6 a.m.), reinforcing the relevance of effective fatigue risk mitigation policies within these less favourable hours of the day. Among the several fatigue countermeasures deeply discussed in Refs. \citep{caldwell2005, caldwell2009}, controlled rest does represent an effective method to mitigate fatigue. As shown quite extensively in the literature \citep{caldwell2012, rosekind1994, neri2002, hartzler2014}, naps have a consistent positive impact to mitigate the adverse effects of sleep loss and/or circadian disruptions. Unfortunately, the Brazilian regulations (RBAC 117) do not allow the use of controlled rests for minimum crew, which is allowed, for instance, in Australia, Bolivia, Canada, China, Europe, Israel, New Zealand, Turkey, and the United Arab Emirates \citep{fsf2018}. Another relevant drawback of not having controlled rest allowed by the regulatory framework is the potentially hazardous occurrence of unintentional naps during flight operations. In fact, a recent Brazilian study \citep{marqueze2017} found a prevalence of 57.8\% of unplanned sleep in a sample of Brazilian pilots, making explicit the need of a regulatory review. It is worth-mentioning, however, that the controlled rest should not be adopted with the intent of increasing flight and/or duty time limitations, but exclusively for fatigue mitigation purposes.\\
The relative fatigue risk as a function of the time of the day is proportional to the inverse of the effectiveness and has a maximum value around 4 a.m. (see the lower panel of Fig. \ref{fig:Fig4}). Our averaged values for the \textit{RFR}, normalized to unit within 18h00 and 23:59, agree qualitatively with the objective measurements of pilot errors in the cockpit \citep{Mello2008}. However, the comparison between these results should be done with caution. Firstly because pilot errors along the time of the day are not exclusively a consequence of fatigue. Secondly because our relative risk estimate is not parametrized as the probability of human errors in a complex aviation environment, but rather by the probability of railroad accidents. In that sense, the pilot errors reported by Mello et al. \citep{Mello2008} are more closely related with cognitive mishaps, besides several other human factor issues not exclusively related with fatigue (see the Limitations Section). Our calculations for the risk exposure, given by the probability of crewing events as a function of the time of the day, are a factor two (14\%) higher than the figures reported by Mello et al. \citep{Mello2008} (7\%) within 0h00 and 05h59, showing a relevant change of the Brazilian Commercial aviation flight schedules from 2005 to 2019.\\
Tailored analyses of the SF inputs related with afternoon naps prior to night shifts, commuting from home to station (and vice-versa) and the advanced bedtime feature of the model were investigated for two low and two high productivity months of 2019 using the responses of the questionnaire. The rosters for the Ids who reported not used to take afternoon naps prior to night shifts, two hours of commuting and not used to advance the bedtime prior to early-starts shifts were run with the standard and the tailored parametrizations, showing relevant group effects ($p<0.001$) for $EM_{C}$, $RM_{C}$ and $FHA_{C}$ in Wilcoxon signed-rank tests for matched pairs (see Table \ref{tab:table5}). The average $FHA_{C}$ increases by 43 to 63\% when discarding the afternoon naps, 14 to 21\% when increasing the commuting from one to two hours and 35 to 54\% when switching off the advanced bedtime criterion, thus showing the high sensitivity of the SF model to these input parameters.\\
Given that the outbreak of the Covid-19 pandemic in Brazil coincided with the new rules prescribed by the RBAC 117 (March of 2020), the impacts of the new regulatory framework are still unknown, motivating the acquisition of more data to shed light on this issue, as scheduled flights and the commercial aviation industry resume their pre-pandemic levels.

\section{Limitations of the study}

The main limitation of this study is related to the model dependency of all the findings and results. Furthermore, the SF inputs and constraints are set in accordance with subjective assessments from the questionnaire, operational experiences and/or educated guessing. For this reason, objective sleep measurements from actigraphy, for instance, would be highly desirable to provide more accurate estimates for the relevant model inputs and criteria. However, these objective measurements are beyond the scope of this work, which is exclusively dedicated to model analyses.\\
Other minor limitations are the non-inclusion of home standby duties and unwind periods when analysing rosters. The home standby events were not taken into account in our model calculations given the uncertainties of the expected amount and quality of sleep. This decision was taken to avoid bias, since the SAFTE-FAST model prevents any sleep event during the entire standby activities. This does not seem very realistic during the night time, as most aircrew workers stay at their homes or at an adequate rest facility. The unwind periods, which encompass the elapsed time from the end of commuting (station to hotel, station to home or station to rest facility) up to the start of the rest period, were also not included. These periods may vary considerable from person to person and include personal needs of hygiene, eating, social activities in preparing for sleep. Such limitations make clear that the fatigue outcomes obtained should be interpreted as lower bounds of fatigue, since some of the crew members might have poor or actually no sleep during home standby duties, as well as relevant unwind episodes at home, hotel or rest facility.\\
Another relevant limitation is related with the extrapolation of the probability of railroad accidents for the aviation scenario. Aviation accidents have a low absolute probability, making it difficult to establish a relationship, for instance, between the SAFTE-FAST effectiveness and human factor accidents with the desirable statistics, as the one obtained in Ref. \citep{rod2020}. Indeed, the investigation of 55 human-factor accidents in aviation \citep{goode2003} demonstrates a relative incidence (accidents proportion per exposure proportion) 5 times higher for duties with 13 hours or more, when compared with duties up to 9 hours.
However, these data do not allow the delineation of statistically relevant relationships between fatigue outcomes and accident risks, since only eleven accidents occurred above 10 hours in duty. In this regard, the comparison of our relative risk ratios as a function of the time of the day with objective measurements of pilot errors in the cockpit \citep{Mello2008} (see the upper panel of Fig. \ref{fig:Fig5}) should be done with caution and under a qualitative approach.
   
\section{Conclusions}
This work adopts a modelling approach to provide a comprehensive statistical analysis of the root causes of fatigue in a robust sample of aircrew rosters of the Brazilian regular aviation, derived from the \textit{Fadigômetro} database. The SAFTE-FAST fatigue outputs and some \textit{productivity metrics} delineate an overall fatigue profile for minimum crew, which show a workload decrease comparing early 2019 and 2020. The rosters are fully characterized by non-linear relationships between the SAFTE-FAST variables of minimum effectiveness (and the maximum equivalent time awake) and the number of night shifts, as well as, the fatigue hazard area and the number of departures and landings within 2 and 6 a.m. (WOCL period), all considered during the critical phases of flight. The 95\% confidence intervals for all the fittings were calculated with the covariance matrix of the fitted parameters and using standard uncertainty propagation techniques. The several structures found for the monthly averaged fatigue hazard areas are consistently interpreted using the distributions of WOCL operations.
The relative fatigue risk increases by 23.3\% (95\% CI, 20.4-26.2\%) increasing the number of night shifts from 1 to 13. Moreover, the relative risk ratios as a function of the time of the day agree qualitatively with pilot errors in the cockpit. On the other hand, the risk exposure found in this work (14\%) is a factor two higher than the figures reported by Mello et al. \citep{Mello2008}. Tailored analyses of some key SAFTE-FAST inputs were done by switching off afternoon naps prior to night shifts, increasing the commuting from home to station (and vice-versa) from 1 to 2 hours and switching off the advanced bedtime criterion of the model, showing significant group effects ($p<0.001 $) for all variables when compared with the standard parametrization. Such finding shows the high sensitivity of the model to these parameters and the need of a deeper investigation to determine more accurately the associated fatigue risk factors. The impact on fatigue caused by the regulatory change with RBAC 117 is still unknown, given the time coincidence with the Covid-19 outbreak in Brazil by mid-March 2020. More studies - preferably aggregating objective sleep measures from actigraphy - are very welcome to provide more stringent constraints to the model inputs and criteria.   

\section{CRediT author statement}
\textbf{Tulio E. Rodrigues:} Conceptualization, Methodology, Validation, Formal analysis, Data Curation, Writing - Original Draft and Project Administration. \textbf{Frida M. Fischer}: Conceptualization, Methodology, Writing - Review \& Editing, Supervision. \textbf{Otaviano Helene}: Methodology, Formal analysis, Writing - Review \& Editing, Supervision. \textbf{Eduardo Antunes}: Conceptualization, Validation, Data Curation, Writing - Review \& Editing. \textbf{Eduardo Furlan}: Validation, Data Curation, Writing - Review \& Editing. \textbf{Eduardo Morteo}: Conceptualization, Validation, Data Curation, Writing - Review \& Editing. \textbf{Alfredo Menquini}: Conceptualization, Validation, Data Curation, Writing - Review \& Editing. \textbf{João Lisboa}: Validation, Data Curation, Writing - Review \& Editing. \textbf{Arnaldo Frank}: Writing - Review \& Editing. \textbf{Alexandre Simões}: Writing - Review \& Editing. \textbf{Karla Papazian}: Data Curation. \textbf{André F. Helene}: Conceptualization, Methodology, Writing - Review \& Editing, Supervision.

\section{Acknowledgment}

We thank the Brazilian Association of Civil Aviation Pilots (ABRAPAC), Gol Aircrew Association (ASAGOL), LATAM Aircrew Association (ATL) and the National Aircrew Union (SNA) for their financial support; Mr. Denys Sene, from IASERA, for the development and support to the roster conversion web-based platform; to Dr. Steven Hursh for the fruitful scientific exchanges and contributions to the manuscript and staff at Institutes for Behavior Resources (IBR) for the SAFTE-FAST customization for the hazard area calculation. We also thank the National Commission of Human Fatigue (CNFH) and Azul Linhas Aéreas Brasileiras for their institutional support in promoting, endorsing and encouraging the study within the aviation community. We finally thank all the crew members who voluntarily agreed to participate in the study.

\section{Supplementary Section} 

\subsection{Most frequent locations of the roster's sample}

\begin{center}
\begin{figure}
\includegraphics[scale=0.37]{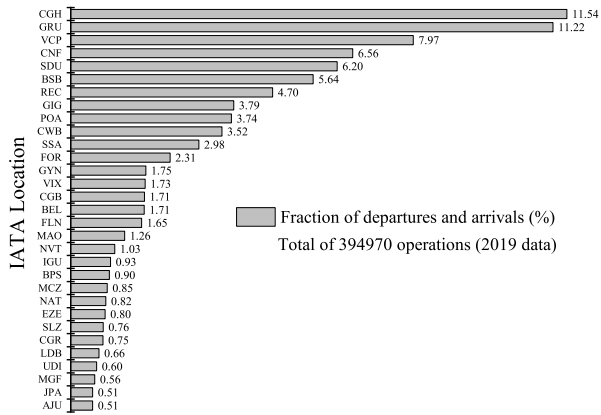}
\caption{\label{fig:Fig6} Percentage of departures and arrivals for the airports with more than 0.5\% of the total operations of 2019.}
\end{figure}
\end{center}

The dominant domestic short-haul characteristic of the roster's sample is clearly shown in Figure ~\ref{fig:Fig6}, which presents the fraction of departures and landings for the airports with more than 0.5\% of all the 394,970 flight operations of 2019. Buenos Aires (EZE), with a frequency around 0.8\%, stands alone as the only foreign destination among the 31 most frequent locations. Congonhas (CGH), Guarulhos (GRU), Campinas (VCP), Confins (CNF), Santos Dumont (SDU) and Brasilia (BSB) altogether comprise 49.1\% of the total flight operations (\textit{Crewing} events only).

\subsection{Epochs for the analyses}

Exact 30 and 15-day epochs were adopted to standardize the extraction of flight schedules, given that fatigue hazard area and duty time have a cumulative character. Also, the number of night shifts, consecutive night shifts, crewing/working events and WOCL operations also depend on the time interval. Even the SAFTE-FAST minimum effectiveness and minimum sleep reservoir depend on the time interval, as longer months have a higher probability for lower scores by chance. Table \ref{tab:table6} shows the 30 and 15-day epochs adopted in this work.    

\begin{table} [b]
\caption{\label{tab:table6}
30 and 15-day epochs adopted for the extraction of flight schedules from the \textit{Fadigômetro} database.}
\begin{ruledtabular}
\begin{tabular}{cccc}

\multirow{2}{*}{Period} & \multirow{2}{*}{Begin date\footnotemark[1]}  &\multirow{2}{*}{End date\footnotemark[1]} & time interval\\
& & & (days)\\ 
\hline
Jan-19& 	1/1/19 0:00&	 1/31/19 0:00 &	30\\
Feb-19& 	1/31/19 0:00&	3/2/19 0:00&	30\\
Mar-19& 	3/2/19 0:00&	 4/1/19 0:00	& 30\\
Apr-19& 	4/1/19 0:00&	 5/1/19 0:00	& 30\\
May-19& 	5/1/19 0:00&	 5/31/19 0:00&	30\\
Jun-19& 	5/31/19 0:00&	6/30/19 0:00&	30\\
Jul-19& 	7/1/19 0:00&	 7/31/19 0:00&	30\\
Aug-19& 	8/1/19 0:00&	 8/31/19 0:00&	30\\
Sep-19& 	8/31/19 0:00&	9/30/19 0:00&	30\\
Oct-19& 	10/1/19 0:00	& 10/31/19 0:00&	 30\\
Nov-19& 	10/31/19 0:00&	11/30/19 0:00&	30\\
Dec-19& 	12/1/19 0:00	& 12/31/19 0:00& 	30\\
Jan-20& 	1/1/20 0:00& 	1/31/20 0:00	& 30\\
Feb-20& 	1/31/20 0:00&	3/1/20 0:00& 	30\\
Mar-19 (1/2)	& 3/1/19 0:00&	3/16/19 0:00	&15\\
Mar-20 (1/2)& 	3/1/20 0:00& 	3/16/20 0:00&	15\\

 \end{tabular}
\end{ruledtabular}
\footnotetext[1]{mm/dd/yy hh:mm}
\end{table}

\subsection{SAFTE-FAST parameters and criteria}
In this section we describe all the parameters and criteria - adopted to mimic behaviours and operational routines before, during and after the working and crewing activities of the Brazilian civil aviation aircrews (Part 121 Passenger Operations) - for the runs with the SAFTE-FAST software version 4.0.3.207.
\subsubsection{SAFTE-FAST input parameters}

For most cases, the start (check-in) and the end (check-out) of the duty periods were captured directly from the rosters. For some rosters format, however, these information were not available and we set 60 minutes prior to the take-off for the check-in and 30 or 45 minutes after the landing for the check-out, depending if the flight sector was domestic or international, respectively. These figures follow the usual practice for scheduled flights in Brazil and also comply with current regulations \cite{Brasil2017, Anac2019}.\\
For the input commuting we set a \textit{standard} 60-minute time interval from home to station, hotel to station and rest facility to station and vice-versa. We also applied an \textit{extended} commuting of 120 minutes from home to station and vice-versa for some of the runs, based upon the responses of the 	questionnaire.\\
For the preparation time, defined as the average time the crew member usually takes to prepare himself for the flight, we set 60 minutes either at home, hotel or rest facility.\\
Owing to the lack of data or reliable information, the unwind time at home, hotel or rest facility was set to zero in our SAFTE-FAST input (see the Limitation Section).\\
Regarding the Auto-Sleep controls of the SAFTE-FAST, we have included both the Auto-Nap and the Advanced bedtime functions. The Auto-Nap function adds automatically an afternoon nap prior to night shifts. The amount of nap depends on the continuous wakefulness period until the last sleep event. For 8 to 10 hours, 10 to 12 hours, 12 to 14 hours or more than 14 hours since the last sleep event, the software adds 60, 90, 120 or 180 minutes of nap, respectively. The Advanced bedtime function allows the software to anticipate the beginning of sleep if an early start would significantly shorten the typical sleep quantity of 8 hours. Both the Auto-Nap and the Advanced bedtime functions can be switched off, as shown for the tailored analyses presented in Table \ref{tab:table5}.
All the sleep metrics adopted in the SAFTE-FAST runs are summarized in Table \ref{tab:table7}.

\begin{table}
\caption{\label{tab:table7}
Sleep metrics adopted in the SAFTE-FAST runs.}
\begin{ruledtabular}
\begin{tabular}{cc}

SAFTE-FAST Auto-Sleep & \multirow{2}{*}{Value or Status}\\
Controls and Parameters & \\
\hline
\multirow{2}{*}{AUTO-NAP function}& ON (standard)\\ 
& OFF (tailored)\\
\multirow{2}{*}{Advanced bedtime function}& ON (standard)\\ 
& OFF (tailored)\\
Normal bedtime & 11 p.m.\\
Minimum Sleep Duration & 60 minutes\\
Maximum Wok Day Sleep & 8 hours\\
Maximum Rest Day Sleep & 9 hours\\
Max Recovery Nap \footnotemark[1] & 210 minutes\\
Awake Zone \footnotemark[2] & 1 to 8 p.m.\\
Sleep quality (home, hotel and rest facility) & Excellent\\
Inflight Sleep & Not included\\

 \end{tabular}
\end{ruledtabular}
\footnotetext[1]{Recovery Nap is automatically added following work duties that end between Normal bedtime and the start of the Awake Zone if the sleep in the past 16 hours is not optimal.}
\footnotetext[2]{The period of the day that the software prevents sleep events, except for afternoon naps prior to night shifts or advanced bedtime events due to early-start shifts.}
\end{table}

\newpage

\bibliography{Manuscript_AAP}

\end{document}